\documentclass[twocolumn,aps,prb,epsf,showpacs]{revtex4}
\usepackage{amsmath}
\usepackage{epsfig}
\usepackage{epsf}
\usepackage{array}

\begin{document}

\title{Benchmarking a semiclassical impurity solver for dynamical-mean-field theory: 
self-energies and magnetic transitions of the single-orbital Hubbard model}
\author{Satoshi Okamoto,$^1$\cite{Email} Andreas Fuhrmann,$^2$ Armin Comanac,$^1$ 
and Andrew J. Millis$^1$}
\affiliation{
$^1$Department of Physics, Columbia University, 538 West 120th Street, New York, 
New York 10027, USA\\
$^2$Physikalisches Institut, Universit{\"a}t Bonn, Nu{\ss}allee 12, D-53115 Bonn, Germany}
\date{\today }

\begin{abstract}
An investigation is presented of the utility of semiclassical approximations 
for solving the quantum-impurity problems arising in the dynamical-mean-field approach 
to the correlated-electron models. 
The method is based on performing a exact numerical integral over the zero-Matsubara-frequency 
component of the spin part of a continuous Hubbard-Stratonovich field, 
along with a spin-field-dependent steepest descents treatment of the charge part. 
We test this method by applying it to one or two site approximations to the single band Hubbard model 
with different band structures, and comparing the results to quantum Monte-Carlo and 
simplified exact diagonalization calculations. 
The resulting electron self-energies, densities of states and magnetic transition temperatures 
show reasonable agreement with the quantum Monte-Carlo simulation over wide parameter ranges, 
suggesting that the semiclassical method is useful for obtaining a reasonable picture of 
the physics in situations where other techniques are too expensive. 
\end{abstract}

\pacs{71.10.Fd, 75.10.-b}

\maketitle

%\draft

\section{Introduction}

Correlated-electron materials, in which the interaction energy is comparable to or larger 
than the electron kinetic energy, are an important topic in materials science. 
In these compounds, standard band-theory is an inadequate representation of the physics. 
The discovery of high-$T_c$ superconductivity in the oxide cuprates\cite{Bednorz86} 
led to greatly increased interest in correlated-electron compounds. 
Many materials have been studied, and many novel properties have been discovered; 
including colossal magnetoresistance, variety of spin, orbital, charge orderings, 
and unconventional superconductivity.\cite{Imada98,Tokura00}
Understanding the novel phenomena and determining the correct electronic phases 
in these materials are challenging tasks, 
and are indispensable for developing the electronic devices exploiting 
the novel properties of the correlated-electron materials. 
However, the rapid increase of Hilbert space with system size limits the utility of 
exact-diagonalizaion (ED) methods, 
while the fermion ``sign problem'' renders Monte-Carlo (MC) approaches ineffective.
 
Dynamical-mean-field theory (DMFT) is a promising approach for treating correlation 
effects.\cite{Metzner89,Georges96} 
The method may be combined with realistic band-structure calculations,  
and has been applied to variety of materials 
(see Refs.~\onlinecite{Anisimov97,Savrasov01,Held01,Dai03} for example).
In DMFT, the momentum dependent electron self-energy is approximated as 
a finite set of functions of frequency, and 
physical lattice problem is mapped into a one or several coupled 
impurity-Anderson models (IAM) with environment determined self-consistently. 
To solve the IAM, a variety of theoretical methods have been applied, including  
analytical approximations such as 
iterated-perturbation theory,\cite{Georges92,Georges93,Kajueter96} 
non-crossing approximation,\cite{Pruschke93} 
projective self-consistent method,\cite{Moeller95} and 
slave-boson \cite{Florens02} for example, 
and numerical methods including 
Quantum Monte-Carlo (QMC),\cite{Jarrell92,Rozenberg92}
exact diagonalization,\cite{Caffarel94} 
numerical-renormalization group,\cite{Sakai94,Bulla99}
and density-matrix-renormalization group.\cite{Garcia04,Nishimoto04} 
Most interesting phenomena can not be addressed by analytical methods. 
Numerical methods, however, are computationally very expensive. 
In order to combine DMFT with a realistic band-structure calculation 
and survey a wide range of parameters, 
it is desirable to develop computationally-cheap and reliable methods. 
Methods which can reproduce the correct electronic phases  
and distinguish the phases near in energy at low temperature are particularly needed. 
Many ideas have been proposed to reduce the computational cost.%
\cite{Potthoff01,Potthoff03,Jeschke04,Zhu04,Savrasov04,Dai04}
These fall mainly into two groups: 
variations of the method simplified by truncating the number of orbitals,\cite{Potthoff01,Potthoff03} 
and perturbative methods involving expansions around a certain solvable limit.%
\cite{Jeschke04,Zhu04,Savrasov04,Dai04}  
The restricted method provides a good approximation to the $T=0$ Mott physics of the Hubbard model,%
\cite{Potthoff01,Potthoff03} however 
because only a small number of states are used, it has difficulty dealing with the thermodynamics 
(as discussed later), and becomes impractical for multiorbital and multisite systems. 
Perturbative methods are not reliable at intermediate coupling. 
Non-perturbative methods which can deal with the thermodynamics are required. 

In this paper, we investigate an alternative method to solve the IAM in DMFT formalism: 
a semiclassical approximation based on the continuous 
Hubbard-Stratonovich transformation.\cite{Hubbard59} 
This method is computationally very cheap (approximately two-orders of magnitude faster than QMC for
single-impurity models, and with a better scaling with system size): 
it may be performed on a commercial PC. 
%for single impurity model, it takes only several minutes to compute one point, 
%while typical computing time for QMC using a similar computer is several hours. 
The semiclassical approximation, in the form used here, was apparently first proposed by Hasegawa,%
\cite{Hasegawa80} who used it to study a ``single-site spin fluctuation theory'' 
which may be viewed as an early, simplified version of dynamical-mean-field theory.
Semiclassical methods were also used by Blawid and Millis\cite{Blawid00} and 
by Pankov, Kotliar and Motome\cite{Pankov02} to study models of electrons coupled to large-mass 
oscillators. 
In this paper, we present an implementation in the context of dynamical-mean-field 
theory, and test its reliability for a fully quantal model problem, namely, the Hubbard model on
a variety of lattices 
by performing detailed comparisons between the semiclassical approximation and QMC. 
We also present a brief comparison a successful ``restricted method'': the two-site DMFT.\cite{Potthoff01} 
The semiclassical approximation is found to be reasonable at high temperature 
and in the strong coupling regime, and gives good results for magnetic transitions. 
In the half-filled square lattice, the N{\'e}el temperature computed by the
semiclassical approximation is very close to that found in QMC. 
In the metallic face-centered cubic (FCC) lattice, ferromagnetic Curie temperatures computed 
by the present method are within a factor of 2 compared with the QMC, and the competition between
ferro- and antiferromagnetism is correctly captured,  
$T=0$ phase boundaries are obtained within the error of $\sim 20~\%$ to QMC, 
The $T=0$ phase boundaries are in the same range as found in two-site DMFT, but 
this latter method gives very poor transition temperatures.  
We also use the semiclassical approximation to compute dynamical properties, in particular, 
the electron self-energy and spectral functions, which are found to be in reasonable agreement 
with QMC results. 

The rest of this paper is organized as follows: 
Section~II formulates the semiclassical approximation. 
In Sec.~III, we use the Gaussian fluctuation approximation 
to estimate the limits of validity of the semiclassical approximation. 
Sections~IV and V compare numerical results for the semiclassical approximation 
to QMC, ED, and Hartree-Fock (HF) results, for 
the single-band Hubbard model on square lattice and face-centered cubic lattice, respectively. 
Section~VI presents an application of the method to the two-impurity cluster dynamical-mean-field 
approximation for the half-filled Hubbard model. 
Comparison of nearest-neighbor spin correlation between the semiclassical approximation 
and QMC are given. 
In Sec.~VII, we discuss the equilibration problem associated with the partitioned phase space 
in the semiclassical approximation and QMC. 
Finally, Sec.~VIII gives a summary and discussion. 

\section{Formulation}

In this section, we derive the semiclassical approximation from 
the continuous Hubbard-Stratonovich (HS) transformation.\cite{Hubbard59} 
We consider a single-orbital repulsive Hubbard model for simplicity: 
$H=H_{band} + H_U$ with 
$H_{band}$ and $H_U$ being single-particle dispersion term and interaction term, respectively.
Generalizations to multiorbital and long-range interaction would be also possible:  
see  e.g., Refs.~\onlinecite{Tyagi75,Hasegawa83,Kakehashi86,Ishihara97} and \onlinecite{Sun02}. 

We take the band-dispersion term to be 
$H_{band}=\sum_{k \sigma} \varepsilon_k c^\dag_{k \sigma} c_{k \sigma}$ 
with three choices of $\varepsilon_k$. \\
{\it Two dimensional square lattice:}
\begin{equation}
\varepsilon^{square}_{\vec k} = -2t (\cos k_x + \cos k_y). 
\label{eq:ek2d}
\end{equation}
{\it Three dimensional FCC lattice:} 
\begin{eqnarray}
\varepsilon^{FCC3}_{\vec k} \!\!\! &=& \!\! 
4t (\cos k_x \cos k_y + \cos k_y \cos k_z + \cos k_z \cos k_x) 
\nonumber \\
&&+2t' (\cos 2 k_x + \cos 2 k_y + \cos 2 k_z), \label{eq:FCC3d}
\end{eqnarray} 
where $t$ and $t'$ are the NN hopping and the second NN hopping, respectively. \\
{\it Infinite-dimensional FCC lattice:} 
Non-interacting density of states is given by 
\begin{equation}
N^{FCC \infty}(\varepsilon)=\frac{\exp \{-(1+\sqrt{2}\varepsilon)/2\}}
{\sqrt{\pi(1+\sqrt{2}\varepsilon)}},
\label{eq:Ulmkedos}
\end{equation}
which diverges at the bottom of the band. 
$N^{FCC \infty}$ was introduced by M{\"u}ller-Hartmann\cite{Hartmann91} 
and studied in detail by Ulmke.\cite{Ulmke98} 
The FCC lattices are of interest because these display a ferromagnetic phase 
in a wide range of doping. 

The interaction term is given by 
\begin{equation}
H_U = U \sum_i n_{i \uparrow} n_{i \downarrow}
\end{equation}
with $U>0$. 

In the DMFT approximation, one first needs to compute the partition function of 
an $N$-site impurity model as 
\begin{equation}
Z = \int {\cal D} [c^\dag, c] \exp\{-({\cal S}_0+{\cal S}_{int})\},
\label{eq:Z}
\end{equation}
with 
\begin{eqnarray}
{\cal S}_0 = -\int_0^\beta \!\! d\tau d\tau' \psi_\sigma^\dag (\tau) {\bf a}_\sigma (\tau-\tau') 
\psi_\sigma (\tau')
\end{eqnarray}
where, 
$\psi = [c_1, \ldots, c_N]^t$ with 
$c_i (c_i^\dag)$ being a Grassmann number corresponding to the electron annihilation 
(creation) operator at site or orbital $i$. 
${\bf a}_\sigma$ is the $N \times N$ matrix Weiss field (inverse of the non-interacting Green function) 
which will be determined self-consistently, and $\beta = 1/T$ is the inverse temperature. 
${\cal S}_{int}$ represents the interaction term specified by the model one considers. 
For the single-band Hubbard model, ${\cal S}_{int} = 
U \! \int \! d \tau \sum _in_{i \uparrow} (\tau) n_{i \downarrow} (\tau)$. 

Next, one computes single-particle interacting Green function ${\bf G}_\sigma$ 
by a functional derivative of ${\rm ln} Z$ with respect to ${\bf a}_\sigma$ as 
\begin{eqnarray}
{\bf G}_\sigma = \frac{\delta {\rm ln} Z}{\delta {\bf a}_\sigma}. 
\label{eq:Gimp}
\end{eqnarray}
The electronic self-energy is obtained by inverting the Dyson equation as 
\begin{equation}
\mbox{\boldmath $\Sigma$}_\sigma = {\bf a}_\sigma - {\bf G}_\sigma^{-1}. 
\label{eq:Dyson}
\end{equation}
Finally, by connecting the impurity Green function $\bf G$ and the local part of the lattice Green 
function, the DMFT equation is closed. 
The self-consistency equation for the $ij$ component of ${\bf G}$ is
\begin{equation}
G_{ij \sigma} (i \omega_n)= 
\int \! \biggl(\frac{dk}{2\pi}\biggr)^d 
\biggl[\frac{\varphi_{ij}(k)}{i \omega_n + \mu - {\bf t}_k - \mbox{\boldmath $\Sigma$}_{k \sigma} 
(i \omega_n)}\biggr]_{ij}, 
\label{eq:Gloc}
\end{equation}
where, $\omega_n$ is the fermionic Matsubara frequency, and 
${\bf t}_k$ and $\mbox{\boldmath $\Sigma$}_k$ are the Fourier transforms of 
the hopping and the self-energy matrices, respectively. 
$\varphi_{ij}(k)$ is a form factor specified by the DMFT method chosen.\cite{Okamoto03} 
The chemical potential $\mu$ is fixed so that ${\bf G}$ gives the correct electron density $n$. 
The local Green function ${\bf G}$ is used to update the Weiss field ${\bf a}_\sigma^{new}$ as 
${\bf a}_\sigma^{new} = {{\bf G}_\sigma}^{-1} + \mbox{\boldmath $\Sigma$}_\sigma$, 
and this process is repeated until the self-consistency ${\bf a}_\sigma={\bf a}_\sigma^{new}$ is obtained. 
The expensive computational task is evaluating the functional integral in Eq.~(\ref{eq:Z}). 

A key point of the present approximation for the single-band Hubbard model 
is introducing the charge field as well as spin. 
Using the following identity: 
\begin{equation}
n_\uparrow n_\downarrow = \frac{1}{4} 
\bigl(\rho^2 - m^2 \bigr), 
\label{eq:identity}
\end{equation}
with $\rho=n_\uparrow+n_\downarrow$ and $m=n_\uparrow-n_\downarrow$, 
we perform the HS transformation 
$\exp (-{\cal S}_{int}) = \int {\cal D} [\varphi, x] \exp (-{\cal S}'_{int})$ 
with the effective interaction 
\begin{eqnarray}
{\cal S}'_{int} \!\! &=& \!\!
\frac{1}{4U} \int_0^\beta \!\! d\tau 
\biggl[ \bigl\{ \varphi^2 (\tau) + x^2 (\tau) \bigr\} \nonumber \\
&& \!\! - 2U \sum_\sigma c^\dag_\sigma (\tau) \{ \varphi(\tau) \sigma_z + i x (\tau) \} c_\sigma (\tau) 
\biggr] ,
\label{eq:Sint}
\end{eqnarray}
where, $\varphi$ and $x$ are the HS fields acting on spin and charge degrees of freedom, respectively, 
and $\sigma_z$ is the $z$ component of the Pauli matrices. 
Now, one can perform Grassmann integral over the fermionic field formally to obtain the partition function 
\begin{equation}
Z=\int \! {\cal D}[\varphi , x] \exp \bigl( -{\cal S}_{eff} \bigr) 
\label{eq:Zfull}
\end{equation}
with the effective action 
\begin{eqnarray}
{\cal S}_{eff} \!\! &=& \!\!
\frac{1}{4U} \int_0^\beta \! d\tau \bigl\{ \varphi^2 (\tau) + x^2 (\tau) \bigr\} \nonumber \\
&& \!\! - {\rm Tr} \, {\rm ln} \Bigl[-a_\sigma 
- \frac{1}{2} \bigl(\varphi \sigma_z + i x \bigr) \Bigr] . 
\label{eq:Seff}
\end{eqnarray}
Here, the trace includes spin as well as Matsubara frequency. 
It is noted that the coupling constant for the charge field $x$ is imaginary. 
This originates from the different sign of quadratic terms for spin and charge 
when decoupling the interaction term. [see Eq.~(\ref{eq:identity})]

\begin{figure}%[tbp]
\begin{center}
\includegraphics[width=0.75\columnwidth,clip]{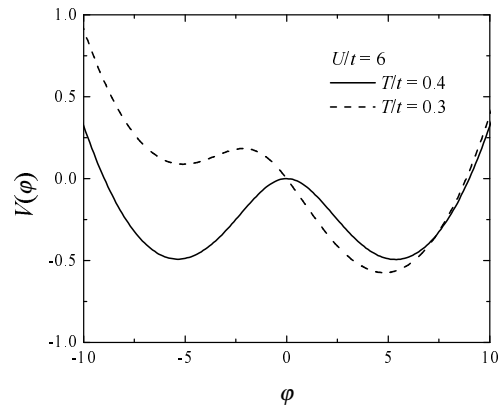}
\end{center}
%\par
%\epsfxsize=0.9\columnwidth \centerline{\epsffile{}}
\caption{Example of the effective potential for $\varphi$, 
$V(\varphi)$ given in Eq.~(\ref{eq:Veff}), 
in the square-lattice half-filled Hubbaed model with $U/t=6$. 
Solid line: $T/t=0.4$ (about 14~\% above $T_N \sim 0.35t$); 
broken line: $T/t=0.3$ (about 14~\% below $T_N$).
At both temperatures, $V$ is far from a simple parabola, and regions far from the local minima 
have appreciable occupation probability. 
}
\label{fig:Veff}
\end{figure}

\begin{figure}%[htbp]
\begin{center}
\includegraphics[width=0.8\columnwidth,clip]{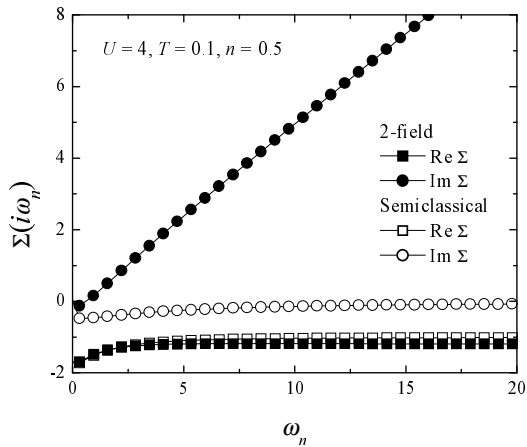}
\end{center}
\caption{Comparison of the electron self-energy computed 
by evaluating the static average integrating over two static fields $\varphi$ and $x$ 
with the action Eq.~(\ref{eq:Sstatic}) (2-field)
and the semiclassical approximation, which is defined in this section, 
for the infinite-dimensional FCC lattice Eq.~(\ref{eq:Ulmkedos}). 
Squares (circles) are real (imaginary) part of the self-energy, and 
filled and open symbols are obtained by 2-field approximation and the semiclassical 
approximation, respectively. 
}
\label{fig:Sigma_bad}
\end{figure}

Exact evaluation of the partition function Eq.~(\ref{eq:Zfull}) is impossible. 
The simplest approximation is to approximate the integrals by the value computed using 
the static parts $\bar \varphi$ and $\bar x$ which extremize the action 
(solutions of the saddle-point equations 
$\partial {\cal S}(\bar \varphi, \bar x)/\partial \bar \varphi = 
\partial {\cal S}(\bar \varphi, \bar x)/\partial \bar x = 0$). 
This is the Hartree-Fock (HF) approximation, 
which is known to give a poor estimates to transition temperatures and self-energies.  
One may correct the HF approximation by including the Gaussian fluctuations in which 
${\cal S}_{eff}$ is expanded around the saddle point 
up to the quadratic order of the fluctuations $\delta \varphi(\tau)$ and $\delta x (\tau)$.  
In this case, $\delta \varphi (\tau)$ and $\delta x (\tau)$ decouple. 
However, Gaussian fluctuation theory is limited to the weak coupling regime. 
As an example, Fig.~\ref{fig:Veff} shows an effective potential $V$ for $\varphi$, 
equivalent to $T {\cal S}_{eff}$, calculated (as explained below) by the semiclassical approximation. 
It is seen that the potential is highly non-parabolic, 
and the variation in $V$ is on the scale of $T$ for reasonable $T$. 
Thus, the Gaussian fluctuation theory is inapplicable. 
Next, one can consider taking $\varphi$ and $x$ only at Matsubara-frequency $\nu_l =0$, 
i.e., static approximation, but evaluating the partition function $Z$ 
as a two-dimensional integral over two static fields (2-field approximation). 
The partition function is expressed as 
$Z_{static}=\int \! d\varphi \, dx \exp(-{\cal S}_{static})$ with
\begin{equation}
{\cal S}_{static} = 
\frac{\beta}{4U} \bigl( \varphi^2 + x^2 \bigr) 
- {\rm Tr} \, {\rm ln} \Bigl[-a_\sigma 
- \frac{1}{2} \bigl(\varphi \sigma_z + i x \bigr) \Bigr].
\label{eq:Sstatic}
\end{equation}
However, this approximation fails 
because the effective action ${\cal S}_{static}$ is complex and 
$\exp (- {\cal S}_{static})$ can not be regarded as a distribution function 
of $\varphi$ and $x$, leading to poor convergence of integrals. 
With some effort, apparently converged integrals can be obtained, 
but the interacting Green function computed by using Eq.~(\ref{eq:Gimp}) with 
$Z$ replaced by $Z_{static}$ does not behave correctly; 
the imaginary part of the self-energy changes sign and 
the spectral-function sum-rule is strongly violated. 
As an example, Fig.~\ref{fig:Sigma_bad} compares self-energies computed by evaluating 
the static average integrating over two static fields $\varphi$ and $x$ 
with the action Eq.~(\ref{eq:Sstatic}) 
to the semiclassical approximation defined shortly. 
While we have obtained the converged solution along the Matsubara-axis, 
imaginary part of the self-energy changes sign and causality is strongly violated, 
so that the analytic continuation to the real axis is impossible. 

\begin{figure}%[htbp]
\begin{center}
\includegraphics[width=0.8\columnwidth,clip]{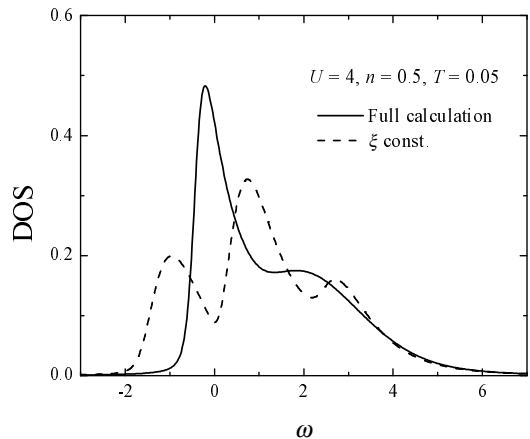}
\end{center}
\caption{Comparison of the many-body density of states computed for the paramagnetic phase of
the infinite-dimensional FCC lattice [Eq.~(\ref{eq:Ulmkedos})] 
by the ``full'' semiclassical approximation (solid line) to that computed using 
additional approximation in which $\xi$ is assumed to be independent of $\varphi$ (broken line). 
The three-peak structure seen in the broken line is unphysical; 
the solid line is in good agreement with QMC (not shown). Note that $T_C \sim 0.073$ for 
these parameters; we have presented data at $T=0.05 \simeq 3T_C/4$ 
(chosen to reveal the three-peak structure clearly), and have artificially suppressed ferromagnetism. 
}
\label{fig:DOS_xi}
\end{figure}

\begin{figure}%[tbp]
\begin{center}
\includegraphics[width=0.75\columnwidth,clip]{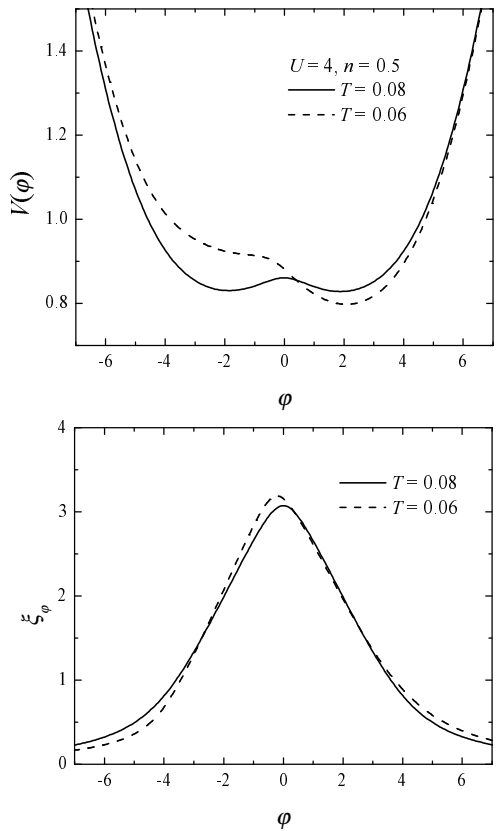}
\end{center}
%\par
%\epsfxsize=0.9\columnwidth \centerline{\epsffile{}}
\caption{Example of the effective potential $V$ (upper panel) and 
charge field $\xi$ (lower panel) as functions of $\varphi$ 
in the metallic Hubbard model on an infinite-dimensional FCC lattice  
[free DOS is given in Eq.~(\ref{eq:Ulmkedos})] with $U=4$ and $n=0.5$. 
Solid lines: $T=0.08$ (about 14~\% above $T_C \sim 0.073$); 
broken lines: $T=0.06$ (about 18~\% below $T_C$). 
The Hartree contribution $(= n U)$ has been subtracted from $\xi$. 
At both temperatures, potential is seen to be highly deviated from simple parabola.
$\varphi$-dependence of $\xi$ indicates strong coupling between the two fields.}
\label{fig:VandXi}
\end{figure}

In order to reduce the above mentioned problems and to take into account 
the fluctuation of both fields to some extent, 
we apply the semiclassical approximation following Hasegawa.\cite{Hasegawa80} 
In this method, we first solve the mean-field equation for the static charge field $\bar x_\varphi$ 
which extremizes ${\cal S}$ at a given value of $\varphi$. 
From $\partial {\cal S}_{eff}/ \partial \bar x |_\varphi = 0$, we obtain the following equation 
$(\xi_\varphi = i \bar x|_\varphi)$, 
\begin{eqnarray}
\xi_\varphi = - U T \, {\rm Tr} \frac{1}{a_\sigma+(\varphi \sigma_z + \xi_\varphi)/2}. 
\label{eq:xi}
\end{eqnarray}
By solving Eq.~(\ref{eq:xi}), one obtains $\xi_\varphi$ as a function of 
$\varphi$. 
In the single-impurity model, this equation has a unique solution, 
thus, Eq.~(\ref{eq:xi}) can be solved easily, for example, via bisection. 
(For multiimpurity models, we have no proof that there is a unique solution, but difficulties have 
not arisen in the case we have studied.) 
Fluctuations of the charge field around this saddle point are expected to be less important than
those of spin field, because of the unique solution and because charge fluctuations are not soft, and 
the difficulties associated with the oscillatory convergence of Eq.~(\ref{eq:Sstatic}) are avoided. 

The $\varphi$-dependence of $\xi$ is crucial. 
If it is neglected, i.e., if $\xi$ is taken to be the value $\xi_{av}$ 
which extremizes $S_{eff}$ after averaging over $\varphi$, 
then one obtains a model equivalent to the Jahn-Teller phonon model studied for example in 
Ref.~\onlinecite{Millis96}. This latter model has different physics from the Hubbard model. 
In particular, the Jahn-Teller model has a spectral function with a characteristic three-peak 
structure quite unlike the spectral function of Hubbard-like models. 
An example of the density of states computed by the approximation, 
$\xi_\varphi \rightarrow \xi_{av}$ is shown in 
Fig.~\ref{fig:DOS_xi} as the broken line. One can clearly observe a three-peak structure. 
The outer two peaks correspond to the occupied (lower band) and unoccupied (upper band) state 
at the occupied distorted site (from potential minima at $\varphi \sim \pm U$), 
and the middle peak corresponds to the unoccupied undistorted state 
(from potential minimum at $\varphi \sim 0$) in the phonon model. 
In the phonon model, the level separation between the lower occupied state and 
the middle unoccupied one has physical meaning as ``polaron binding,'' which does not exist 
in the original Hubbard model. 
The spectral function computed from the ``full'' semiclassical method (see below) 
is shown as the solid line in Fig.~\ref{fig:DOS_xi}, and as will be seen below is in much better 
agreement with QMC data.  

Now that the saddle point of the effective action is determined with respect 
to one of two variables, $\xi$, 
an effective potential for remaining variable $\varphi$ is written as 
\begin{equation}
V(\varphi) = \frac{1}{4U} (\varphi^2 - \xi_\varphi^2)
- T \, {\rm Tr} \, {\rm ln} \Bigl[-a_\sigma 
- \frac{1}{2} \bigl(\varphi \sigma_z + \xi_\varphi \bigr) \Bigr]. 
\label{eq:Veff}
\end{equation}
With this effective potential, the partition function is approximated as 
\begin{eqnarray}
Z_{approx}=\int_{-\infty}^\infty d \varphi \exp\{-\beta V(\varphi)\}. 
\label{eq:Zapprox}
\end{eqnarray}
There remains only one variable $\varphi$, which is purely real. 
Numerical integrals can be performed without difficulty. 
Figure~\ref{fig:VandXi} shows the example of $V(\varphi)$ and $\xi_\varphi$ 
calculated by ``full'' semiclassical approximation for a non-integer filling. 
It is seen again that $V$ is highly non-parabolic. 
Furthermore, $\xi_\varphi$ depends on $\varphi$ indicating the strong coupling between 
spin- and charge-fields. 

Now the approximate partition function $Z_{approx}$ is obtained, 
one can obtain physical quantity from the functional derivative form 
following the DMFT procedure. 
The impurity Green function is
\begin{eqnarray}
G_\sigma = \frac{\delta {\rm ln} Z_{approx}}{\delta a_\sigma}
=\bigg\langle \frac{1}{a_\sigma+(\varphi \sigma_z + \xi_\varphi)/2} \bigg\rangle,
\label{eq:Gfull}
\end{eqnarray}
where $\langle \ldots \rangle$ stands for 
$\int \! d\varphi \exp\{-\beta V(\varphi)\} \ldots / Z_{approx}$. 
Thus, Eqs.~(\ref{eq:xi})--(\ref{eq:Gfull}) with the DMFT self-consistency equation 
construct the present semiclassical approximation. 

It may be useful to mention the correspondence between the present theory and 
the single-site spin-fluctuation theory of Hasegawa.\cite{Hasegawa80} 
In Ref.~\onlinecite{Hasegawa80}, Hasegawa introduced the same HS transformation using 
spin- and charge-fields, and applied saddle-point approximation against the charge-field. 
Thus, up to this point, two theories are equivalent, but instead of Eq.~(\ref{eq:Gfull}) and
DMFT self-consistency equation, Hasegawa used an ad-hoc procedure involving 
computation of averages of the magnetic moment and square of the magnetic moment. 

We emphasize, however, that the semiclassical approximation is not exact. 
In weak coupling, it leads to a scattering rate $\propto T$, rather than $\propto T^2/E_F$, 
and does not give a mass renormalization; 
in strong coupling, while it gives correct Mott insulator behavior, 
i.e., divergence of the on-site self-energy $\propto 1/(i \omega_n)$, 
the self-energy at small $\omega$ remains finite at $T \rightarrow 0$ 
even in a paramagnetic metallic phase as long as the effective potential $V(\varphi)$ 
has more than one degenerate minimum, 
implying incorrect non-Fermi liquid behavior. 
This originates from the neglect of the quantum fluctuation of HS fields. 
Including the quantum fluctuation to recover the correct Fermi-liquid behavior 
is not easy.\cite{Attias97,Pankov02,Blawid03} 
Although, it is not exact, it will be shown below that the semiclassical method gives good 
estimates of the important physical quantities. 

\section{Weak Coupling Gaussian Approximation: Comparison of Classical  and Exact Results}

This section uses the Gaussian approximation to the paramagnetic phase
of the functional integral to gain insight into the limits of validity of
the classical approximation.  The advantage of the Gaussian approximation is that
any desired quantity can be computed straightforwardly, permitting a detailed
comparison of the results of the  classical approximation to the fully quantal calculation.  
We consider the mean square amplitude of the spin Hubbard-Stratonovich
field, $\left<\varphi^2\right>$, and also the Matsubara-axis 
electron self-energy, $\Sigma(i\omega_n)$.  

In the Gaussian approximation one determines the values 
${\bar \varphi}=UT \, {\rm Tr}_{\omega_n,\sigma}\sigma_z\left[a_0
+\frac{1}{2}({\bar \varphi}\sigma_z+ i \bar x)\right]^{-1}$ and
$\bar x= i UT \, {\rm Tr}_{\omega_n,\sigma}\left[a_0
+\frac{1}{2}({\bar \varphi}\sigma_z+ i \bar x)\right]^{-1}$ 
which extremize the argument of the exponential, and then
expand the argument of the exponential to second order in the deviations
of the fields from their extremal values:
\begin{eqnarray}
Z \rightarrow {\bar Z} \int {\cal D} [\varphi \, x]
\exp \Bigl[-S^{(2)}_{\varphi}-S^{(2)}_{x} \Bigr], 
\end{eqnarray}
where $\bar Z$ is contributions from the saddle points, and 
\begin{eqnarray}
S^{(2)}_{\lambda} \!\!\! &=& \!\!\! \int \! d\tau_1 d\tau_2 \, \lambda(\tau_1)  
\chi_{\lambda}^{-1}(\tau_1-\tau_2) \lambda(\tau_2), \\
\chi_{\varphi/x}^{-1} \!\!\! &=& \!\!\! \frac{1}{4U} \, \delta(\tau_1-\tau_2) \nonumber \\
&& \hspace{-1.5em} \pm \frac{1}{8} {\rm Tr} \Bigr[\mbox{\boldmath $\alpha$}_{\varphi/x} 
{\bf G}_0 (\tau_1-\tau_2)
\mbox{\boldmath $\alpha$}_{\varphi/x}
{\bf G}_0(\tau_2-\tau_1) \Bigl],  \\
{\bf G}_0(\tau) \!\!\! &=& \!\!\! \Bigr[ a_0(\tau){\bf 1}+\frac{1}{2}({\bar \varphi} 
\mbox{\boldmath $\sigma$}_{\!\! z}+ \bar x {\bf 1})\delta(\tau)\Bigl]^{-1},
\label{eq:g0}
\end{eqnarray}
with 
$\mbox{\boldmath $\alpha$}_{\varphi} \! = \! \mbox{\boldmath $\sigma$}_{\!\! z}$ and 
$\mbox{\boldmath $\alpha$}_{x} \! = \! \mbox{\boldmath $1$}$, these are $2\times2$ 
matrices acting on spin space. 
In the mean-field approximation, a critical $U$ exists; for $U<U_c$, $\bar \varphi = 0$, and 
for $U>U_c$, $\bar \varphi \ne 0$. (This transition is removed by fluctuations.)
For $U \agt U_c$, spin fluctuations are very soft. 

We focus here on the $\varphi$ integral, and we consider the small-$U$ regime in which 
$\bar \varphi=0$, and in the paramagnetic phase, the $x$ and $\varphi$ integrals decouple. 
The mean square value of the fluctuations of $\varphi$, $\left< \varphi^2 \right>$ 
(obtained by performing the Gaussian integral over all Matsubara components of $\varphi$)
and the classical approximation $\left< \varphi^2 \right>_{class}$ 
(obtained by performing the integral only over zero Matsubara components of $\varphi$)
are
\begin{eqnarray}
\left< \varphi^2 \right> \!\! &=& \!\! -T\sum_{l}\frac{U^2 \chi_0(i\nu_l)}{1+U \chi_0(i\nu_l)},\\
\left< \varphi^2 \right>_{class} \!\! &=& \!\! -T\frac{U^2 \chi_0(0)}{1+U \chi_0(0)}, 
\end{eqnarray}
where $\chi_0$ is an irreducible susceptibility given by 
\begin{equation}
\chi_0(i \nu_l)= 2 T \sum_n a_0^{-1}(i \omega_n+i\nu_l) \, a_0^{-1}(i\omega_n),
\end{equation}
with $\nu_l$ being a bosonic Matsubara-frequency. 
Finally, the leading perturbative contribution to the electron self-energy $\Sigma$ and 
its classical approximation $\Sigma_{class}$ are given by 
\begin{eqnarray}
\Sigma(i\omega_n) \!\! &=& \!\! \frac{1}{4} \, T \sum_l \chi(i\nu_l) \, 
a_0^{-1}(i\omega_n-i\nu_l), \\
\Sigma_{class} (i\omega_n) \!\! &=& \!\! \frac{1}{4} \, T \chi(0) \, a_0^{-1}(i\omega_n), 
\end{eqnarray}
with the susceptibility 
\begin{equation}
\chi(i\nu_l) = \frac{U^2 \chi_0(i\nu_l)}{1+ U \chi_0(i\nu_l)}. 
\end{equation}

We now present results obtained by applying the Gaussian approximation 
to the paramagnetic phase of the infinite-dimensional FCC lattice 
[DOS is given by Eq.~(\ref{eq:Ulmkedos})]. 
In this section, energy is scaled by the variance of DOS (=1). 
The lower band edge is at $\varepsilon_{min}=-1/\sqrt{2}\approx-0.71$ 
and the square root divergence of the density of states means that for 
small fillings the density of states is very close to the bottom of the 
band. In the non-interacting limit the chemical potential is 
$\mu_{n=0.5}\approx-0.6335$ ($\mu-\varepsilon_{min}  \approx 0.078$) 
and $\mu_{n=0.75}=-0.5352$ ($\mu-\varepsilon_{min} \approx .175$).  
The square root divergences mean that the critical interaction strengths 
beyond which the mean field solution becomes unstable are
themselves temperature dependent, and are small.

\begin{figure}%[tbp]
\begin{center}
\includegraphics[width=0.8\columnwidth,clip]{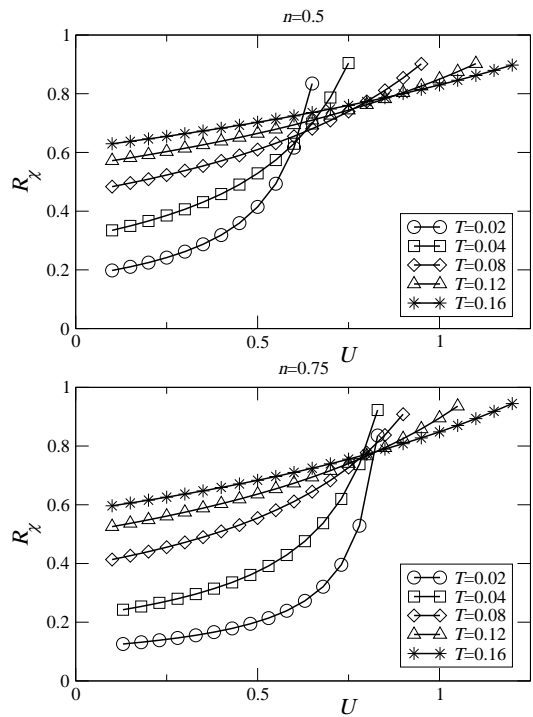}
\caption{Ratio of the classical approximation to the 
full (Gaussian) mean square fluctuation of spin Hubbard-Stratonovich field
$R_\chi=\left<\varphi^2\right>_{class}/\left<\varphi^2\right>$  
for infinite-dimensional FCC lattice at densities $n=0.5$ (upper panel) and $n=0.75$ (lower panel). 
The curves cross because of the temperature dependence of the critical interaction 
strength (evident from the $U$ at which the ratio approaches unity).}
\label{fig:Chi_ratio_ulm}
\end{center}
\end{figure}

\begin{figure}%[tbp]
\begin{center}
\includegraphics[width=0.95\columnwidth,clip]{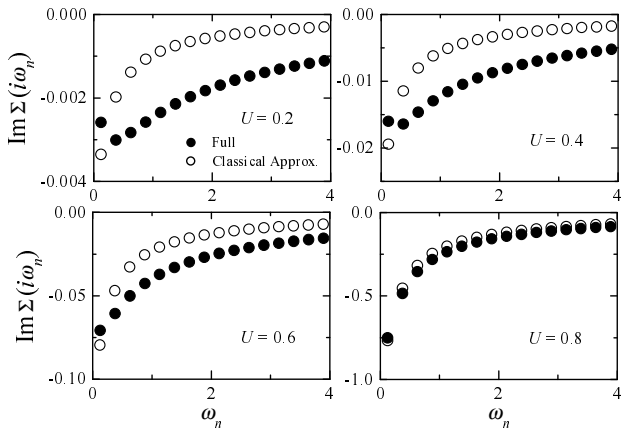}
\caption{Frequency dependence of full (Gaussian) self-energy 
(filled circles) and classical approximation (open circles), for infinite-dimensional FCC lattice
with  $n=0.75$, $T=0.04$ and $U$ values indicated.}
\label{fig:Sigma_ulm_Udep}
\end{center}
\end{figure}

Figure \ref{fig:Chi_ratio_ulm} gives the comparison of the full and Gaussian fluctuation
approximation to the mean square fluctuation of the spin Hubbard-Stratonovich field
for the infinite-dimensional FCC lattice model with $n=0.5$ and $n=0.75$. 
One sees that the classical approximation captures most of the fluctuations either for $U$ near 
the critical value or for temperatures of the order of the bandwidth.
The panels of Fig.~\ref{fig:Sigma_ulm_Udep} show the full self-energy and the classical
approximation for different interaction values, for $n=0.75$. 
For small $U$, the semiclassical approximation fails, but 
as $U$ approaches the critical value (here approximately $0.84$) the self-energy
becomes well represented by its classical approximation

\begin{figure}%[tbp]
\begin{center}
\includegraphics[width=0.95\columnwidth,clip]{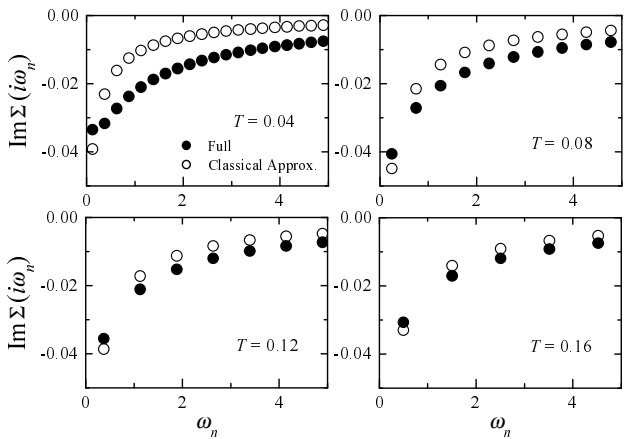}
\caption{Frequency dependence of imaginary part 
of full (Gaussian) self-energy (filled circles) and classical
approximation (open circles) for $n=0.75$, $U=0.5$ and $T$ values indicated.}
\label{fig:Sigma_ulm_Tdep}
\end{center}
\end{figure}

The panels of Fig.~\ref{fig:Sigma_ulm_Tdep} show the 
dependence of the self-energy on temperature, at a fixed moderate
interaction strength (about 3/4 of the critical value).  We see that at
all temperatures studied, the classical approximation provides a reasonable estimate
of the low frequency self-energy. At
temperatures less than about half of the band width, the  classical
approximation grossly underestimates the high frequency part of the
self-energy, but by $T=0.08$ it is within about $30~\%$ of the exact
value, and for higher temperatures or for interactions close to the critical
value the approximation is quite good.

\section{Comparison of semiclassical approximation to QMC: Two-dimensional square lattice}

The Hubbard model on a bipartite lattice is known to exhibit an antiferromagnetic 
N{\'e}el ordering at half-filling and finite $U$. 
In this section, we apply the semiclassical approximation to the Hubbard model with 
a square lattice [non-interacting electron dispersion is given by Eq.~(\ref{eq:ek2d})], 
and investigate the self-energy, density of states and magnetic transition temperature. 
Note that the finite $T_N$ in two-dimensional square lattice is an artifact 
arising from the neglect of the low-lying spin-wave excitation which DMFT method can not capture. 

\begin{figure}%[htbp]
\begin{center}
\includegraphics[width=0.8\columnwidth,clip]{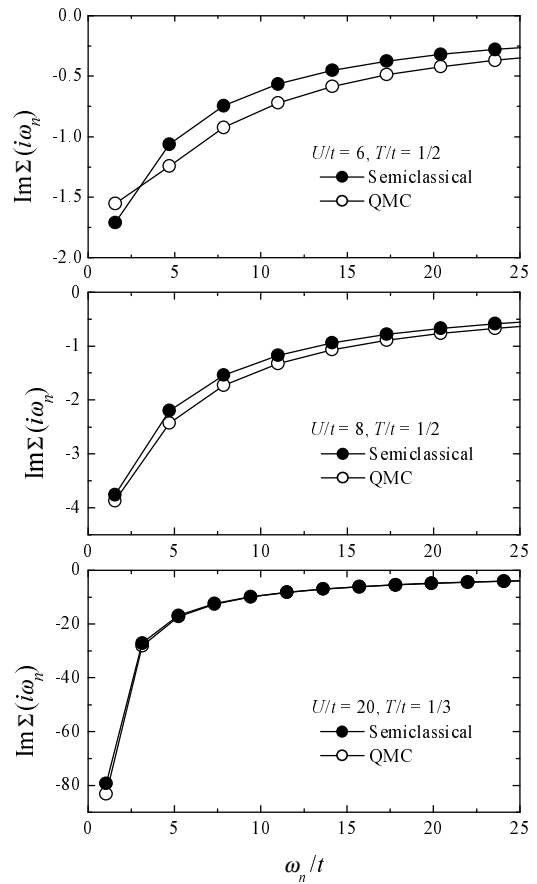}
\end{center}
\caption{Self-energies for a square-lattice half-filled Hubbard model 
computed from the semiclassical approximation (filled circles) and 
QMC (open circles) for interactions and temperatures indicated.}
\label{fig:Sigma_square}
\end{figure}

These results are compared with QMC performed on computing facilities at Universit{\"a}t Bonn 
using the Harsch-Fye algorithm.\cite{Georges96} 
Typically, 48 time slices were used, and 2--20$\times 10^6$ MC configurations were recorded. 
Occasional runs with more time slices or more MC configurations were made to verify error. 
Between 5 and (for the largest $U$) 40 iterations were needed for convergence of the DMFT loops. 
The N{\'e}el temperatures were determined from computations of the staggered magnetization. 
For $U \agt 8t$ and $T$ in or near the magnetic phase boundary, global update techniques 
(to be described elsewhere) were needed to ensure equilibration of the MC calculation. 
Error bars are smaller than the symbols shown. 

As noted in the previous section, it is expected that the semiclassical approximation
becomes accurate in the strong-coupling regime and high temperature. 
In order to confirm this expectation beyond the harmonic (Gaussian fluctuation) approximation, 
we first calculate the electronic self-energies in the paramagnetic state and compare 
them with the QMC results. Recall that for this model the full band-width is $8t$. 
In Fig.~\ref{fig:Sigma_square}, shown are 
the self-energies at $U/t=6$ and $T/t=1/2$ (upper panel), 
$U/t=8$ and $T/t=1/2$ (middle panel), and $U/t=20$ and $T/t=1/3$ (lower panel) 
computed by the semiclassical approximation and QMC as functions of imaginary frequency. 
With increase of $U$, the self-energy at low frequency increases, and for $U > U_c$, diverges 
indicating the Mott metal-insulator transition. 
This behavior is well reproduced by the semiclassical approximation, 
and the agreement with the QMC is remarkable, particularly in the strong coupling regime. 

\begin{figure}%[htbp]
\begin{center}
\includegraphics[width=1\columnwidth,clip]{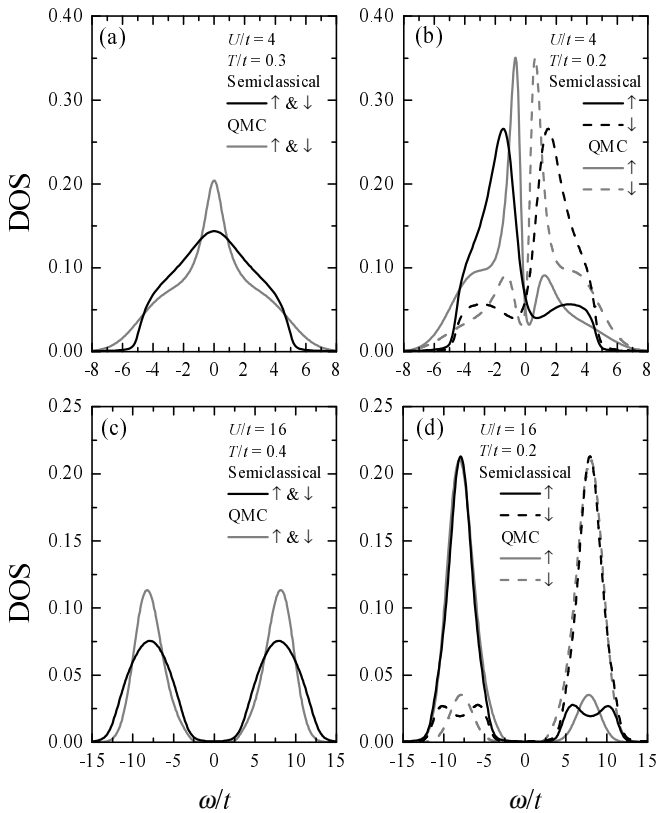}
\end{center}
\caption{Density of states of square-lattice half-filled Hubbard model computed using semiclassical 
method (heavy lines) and QMC (light lines) for paramagnetic [panels (a) and (c)] and antiferromagnetic 
[panels (b) and (d), $T \sim 0.8 T_N$] 
phases at interaction and temperatures indicated. 
In the antiferromagnetic phase, majority spin density of states is shown by solid line, 
minority spin by broken line. 
}
\label{fig:DOS_square}
\end{figure}

Figure~\ref{fig:DOS_square} compares the semiclassical results for a real-frequency quantity, 
the density of states, to results obtained by maximum entropy analytical continuation of QMC data. 
This is a rather stringent rest of the method, and agreement is seen to be reasonably good. 
In the weak coupling, paramagnetic phase [Fig.~\ref{fig:DOS_square}~(a)], 
the semiclassical approximation underestimates the $\omega=0$ peak 
(because $\omega=0$ scattering rate is overestimated) and 
underestimates the weight in the wings (because the high frequency scattering rate is 
underestimated, see the upper panel of Fig.~\ref{fig:Sigma_square}). 
Below $T_N$, the magnetization is slightly overestimated by the semiclassical approximation 
(magnetization $\langle m \rangle \sim 0.54$ for the semiclassical approximation 
and  $\langle m \rangle \sim0.4$ for QMC at $U/t=4$ and $T/t=0.2$) 
as can be seen from the slightly larger gap.  
DOS is sharper because the self-energy at high-frequency is underestimated as in the
paramagnetic state. 
In the strong coupling regime, agreement between the semiclassical approximation and QMC 
is quite good as can be expected from the self-energy 
(see the lower panel of Fig.~\ref{fig:Sigma_square}).

\begin{figure}%[htbp]
\begin{center}
\includegraphics[width=0.8\columnwidth,clip]{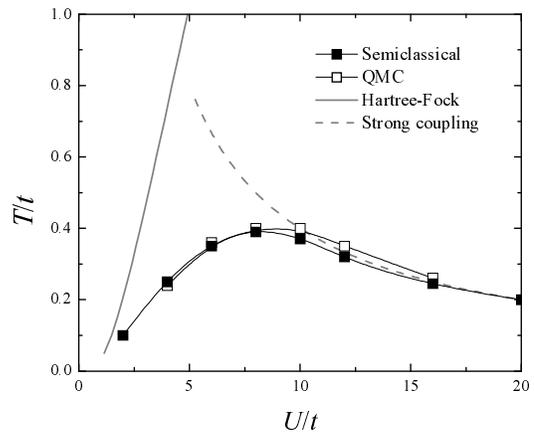}
\end{center}
%\par
%\epsfxsize=0.85\columnwidth \centerline{\epsffile{spectra.eps}}
\caption{N{\'e}el temperature of square-lattice half-filled Hubbard model 
as functions of $U$ computed from semiclassical approximation (filled symbols) 
and QMC (open symbols), HF (light solid line) and large $U$ limit ($T_N = 4t^2/U$)
(light broken line).}
\label{fig:TN_square}
\end{figure}

Figure~\ref{fig:TN_square} summarizes results for the magnetic phase diagram 
obtained from the semiclassical approximation, QMC, the HF approximation 
and a strong-coupling expansion. The semiclassical approximation is seen to give remarkably good 
results over the whole phase diagram. 
In the weak-coupling limit, the semiclassical approximation gives correct behavior: 
$T_N$ asymptotes to the result of HF in the limit of $U \rightarrow 0$. 
This is natural because the two approximations becomes 
identical in the $U \rightarrow 0$, $T \rightarrow 0$ limit. 
At finite but small $U$ region, the reduction of $T_N$ from HF is found to be quite large, 
because the finite self-energy reduces nesting effect substantially. 
The present approximation gives correct behavior in the strong-coupling regime; 
$T_N$ asymptotes to $4t^2/U$, the mean-field result of the strong coupling expansion, 
and, in particular, is almost identical to the QMC results. 
It should be noted that at large $U$ and low $T$, the QMC requires a very large amount of 
Monte-Carlo sampling to reach equilibrium, whereas the semiclassical method is numerically very cheap. 
Even for $U/t=16$ and $T/t \sim 0.2$ (antiferromagnetic phase), 
it takes less than 5~mins. on a commercial PC to compute one point in the present method, 
while it takes $\sim 6$~hrs. by QMC on a similar computer for the same parameters. 

\section{Comparison of semiclassical approximation to QMC: FCC lattice}

In this section, we apply the semiclassical approximation to the single-band Hubbard model on 
the FCC lattice in infinite- and three-dimensions, and compare the results to QMC data of 
Ulmke,\cite{Ulmke98} and to the ``two-site'' approximation of Potthoff.\cite{Potthoff01}
We focus on the filling dependence of the magnetism in these models, 
in which the charge fluctuation plays an important role. 
At non-integer filling, these models order ferromagnetically because of the large DOS 
near the bottom of the band, 
in contrast to the antiferromagnetic ordering found in half-filling square lattice. 
In addition, near half-filling, 
the three-dimensional model is found to show an additional competing order: 
a ``layer-type antiferromagnetic'' state. 
We examine whether the semiclassical approximation capture this behavior. 
In this section, we use the variance of DOS $v$ as an unit of energy.  

\begin{figure}%[htbp]
\begin{center}
\includegraphics[width=0.8\columnwidth,clip]{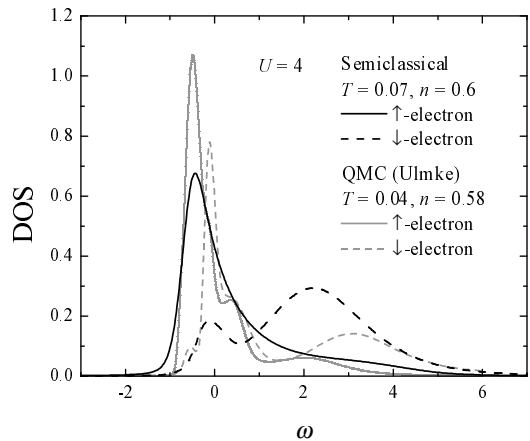}
\end{center}
\caption{Comparison of the density of states of infinite-dimensional FCC Hubbard model 
with $U=4$ and temperatures chosen so that $\langle m \rangle =0.4$ in each case. 
Heavy lines: results of the semiclassical approximation at $n=0.6$ and $T=0.07$ 
($\sim 13~\%$ below $T_C \sim 0.08$). 
Light lines: results of QMC at $n=0.58$ and $T=0.04$ ($\sim 20~\%$ below $T_C \sim 0.05$)
taken from Ref.~\onlinecite{Ulmke98}. 
Solid and broken lines are for the majority and minority spin.}
\label{fig:DOS_fcc}
\end{figure}

First, we apply the semiclassical approximation to 
the infinite-dimensional FCC lattice with DOS given in Eq.~(\ref{eq:Ulmkedos}), $v=1$. 
The density of states above the Curie temperature is essentially the same as seen as 
the solid line in Fig.~\ref{fig:DOS_xi}, and 
consists of two structures: one peak just below the Fermi level $\omega=0$ and 
shoulder at $\omega \sim nU$. 
These correspond to the lower- and upper-Hubbard bands. 
(Note that the result shown in Fig.~\ref{fig:DOS_xi} is computed at $T<T_C$, 
but in the paramagnetic phase, by suppressing the magnetic transition.) 
The separation between the lower- and upper-Hubbard band is somewhat reduced 
from the bare value of $U$ (by about a factor of mean occupation number $n$), 
possibly because the neglect of the fluctuation of a charge-field. 
The two structures continuously evolve into the majority-spin and minority-spin bands 
below $T_C$ as shown in Fig.~\ref{fig:DOS_fcc}. 

Figure~\ref{fig:DOS_fcc} also presents the DOS from QMC. 
(Temperature and density are chosen so that the magnetization in the two calculations are 
similar.)
In QMC results, we observe a sharp peak below Fermi level for the majority spin band. 
For the minority spin band, there appears a sharp peak near the Fermi level and  
a broad hump at $\omega \sim 3$. 
The former originates from the fact that the system is not fully polarized, and 
the latter corresponds to the upper-Hubbard band. 
As noted above, the upper-Hubbard band (broad hump around $\omega=3$ in the minority-spin band) 
appears slightly low in energy in the semiclassical approximation. 
Except for this discrepancy, the overall structure of DOS is well reproduced 
by the semiclassical approximation. 
In particular, the occupied bands below $\omega=0$ show reasonable agreement. 

\begin{figure}%[tbp]
\begin{center}
\includegraphics[width=0.8\columnwidth,clip]{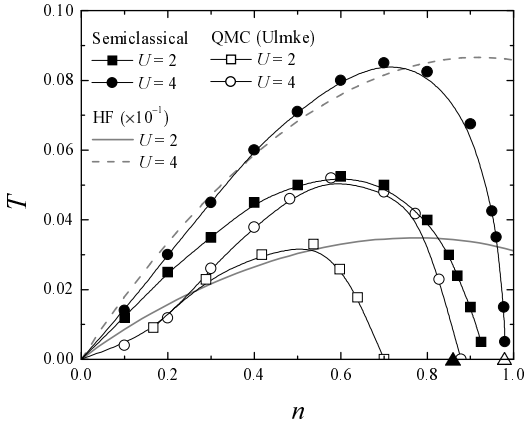}
\end{center}
\caption{Ferromagnetic Curie temperature of single-band Hubbard model on an 
infinite-dimensional FCC lattice as function of electron density for $U=2$ (squares)
and 4 (circles). 
Filled symbols: semiclassical approximation; 
open symbols: QMC (Ref.~\onlinecite{Ulmke98}).
Curie temperatures ($\times 10^{-1}$) computed by HF approximation 
to the same model with $U=2(4)$ are shown as a light solid (broken) line.
Two-site DMFT results for the transition temperatures are not shown, 
but representative two-site results for $n=0.5$ are $T_C \sim 0.17 (0.26)$ for $U=2 (4)$. 
Phase boundary at $T=0$ for $U=2(4)$ computed by the two-site DMFT is 
shown as a filled (open) triangle. }
\label{fig:TC_fcc_inf}
\end{figure}

In Fig.~\ref{fig:TC_fcc_inf}, 
ferromagnetic Curie temperatures $T_C$ for different values of $U$ are shown 
as functions of electron density. 
Filled symbols are the results obtained using the semiclassical approximation. 
For comparison, results of QMC\cite{Ulmke98} and HF are shown as open symbols and 
light lines, respectively (Note that the HF $T_C$ shown are reduced by a factor of 10). 
The semiclassical approximation overestimates $T_C$ than QMC 
by a factor of $\sim 50~\%$ over a wide range of density. 
(Of course, the difference near the critical density of QMC is much larger.) 
It is seen that HF approximation is very poor in all parameter regimes, 
highly overestimating the ferromagnetic Curie temperature. 
The critical density $n_c$ is also overestimated, 
at $n_c \sim 1.2$ at $U=2$ and $n_c \sim 1.5$ at $U=4$. 
The higher $T_C$ found in the semiclassical approximation may be due to the neglect of 
the quantum fluctuation of the HS fields. 
Reduction of the transition temperature due to 
the quantum fluctuation can be seen in the context of electron-phonon coupling 
in Ref.~\onlinecite{Blawid01}. 
However, good agreement between the semiclassical approximation and QMC indicates that 
the thermal part of the fluctuation dominates the electron self-energy in the wide region 
and, thus, the magnetic transition. 
Critical densities $n_c$ where the ferromagnetism disappears at $T=0$ are found to be 
slightly higher in the semiclassical approximation than in QMC, 
$n_c \sim 0.88$ for $U=2$ and $n_c \sim 0.97$ for $U=4$, while 
QMC gives $n_c \sim 0.7$ for $U=2$ and $n_c \sim 0.88$ for $U=4$. 
However, in the light of the simplification of the present approximation, 
the agreement with QMC within 20~\% error is remarkable. 

We applied the two-site DMFT\cite{Potthoff01} to the same model 
to investigate the magnetic phase diagram. 
The critical densities $n_c$ obtained by two-site DMFT are found to be very similar to 
those by the semiclassical approximation. 
However, Curie temperatures are found to be overestimated by a factor of $\sim 6$ compared
with QMC, and by a factor of $\sim 3$--$4$ compared with the semiclassical approximation. 
In the two-site DMFT, $T_C$ at $n=0.5$ is found to be $\sim 0.17$ for $U=2$, and 
$\sim 0.26$ for $U=4$.  
The overestimate arises because, in the two-site DMFT, the IAM is composed of only two sites, 
one is correlated (impurity) and one is non-correlated (bath). 
The energy difference between these sites is generically large, 
so the (spin) entropy is underestimated. 
The two-site DMFT also yields reentrant behavior near the critical concentration, 
which is not observed in the semiclassical approximation and QMC. 
Adding more sites in the ``bath'' part of IAM would remedy these behaviors, 
but at drastically increased computational expense. 

Finally, we investigate the magnetic instability in the more realistic  
three-dimensional FCC lattice whose free band-dispersion is given in Eq.~(\ref{eq:FCC3d}). 
Following Ref.~\onlinecite{Ulmke98}, 
we take $t'= t/4$ and choose the variance of the DOS  $v = \sqrt{12 t^2 + 6 {t'}^2}$ 
as the energy unit, thus $t \approx 0.284$ and $t' \approx 0.071$. 
With this parameter set, the bottom of the band is given by 
$\varepsilon_{min} = -4t+2t' \approx - 0.994$. 
There is no divergence in the free DOS in contrast to the infinite-dimensional case. 

\begin{figure}%[tbp]
\begin{center}
\includegraphics[width=0.8\columnwidth,clip]{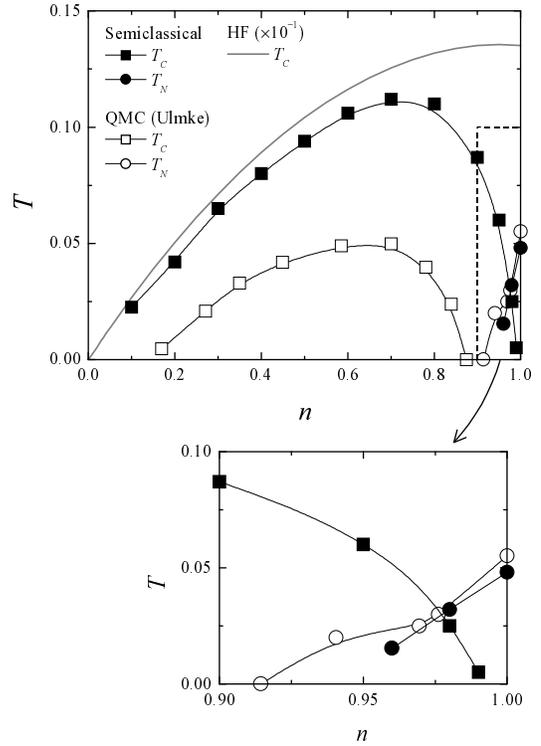}
\end{center}
\caption{Upper panel: Phase diagram of single-band Hubbard model on 
a three-dimensional FCC lattice as a function of electron density and temperature
for interaction $U=6$ and variance of the density of states $v=1$. 
Squares and circles are Curie temperature and N{\'e}el temperature 
for layer-type antiferromagnetic state, respectively. 
Filled symbols: results of the semiclassical approximation; 
open symbols: QMC results taken from Ref.~\onlinecite{Ulmke98}.
Curie temperature ($\times 10^{-1}$) computed by HF approximation 
to the same model is shown as a light solid line.
Lower panel: Expansion of the phase diagram, showing the region near $n=1$ 
enclosed by a broken line in the upper panel. 
The semiclassical results for $T_C$ and $T_N$ indicate the instability of 
ferromagnetic and antiferromagnetic states, 
respectively, towards the paramagnetic state. 
}
\label{fig:TC_fcc_3d}
\end{figure}

The magnetic phase diagram of the three-dimensional FCC lattice 
computed by the semiclassical approximation is shown in Fig.~\ref{fig:TC_fcc_3d}. 
For comparison, QMC\cite{Ulmke98} and HF results are also plotted. 
It is seen that the semiclassical-approximation results for $T_C$ are 
about twice higher than QMC, while HF results are about 20 times higher than QMC. 
Similar to the infinite-dimensional case, higher critical density $n_c$ 
is overestimated by $\sim20~\%$. 
QMC calculations are argued to indicate that there exists a lower critical density 
$n_c' \sim 0.15$ below which the ferromagnetism disappears.\cite{Ulmke98} 
This behavior is ascribed to the absence of the divergence at the bottom of the bare DOS. 
In the semiclassical approximation, 
$T_C$ seems to become zero proportional to the electron density similarly to  
the HF approximation. In the latter, clearly $n_c'=0$. 
This might be because the semiclassical approximation reduces to the HF 
at weak coupling (in this case small density) and low temperature limit. 
However, it is very difficult to judge if $n_c'=0$ or not in the present accuracy for 
the semiclassical approximation. 

Differently from the infinite-dimensional FCC lattice, 
QMC calculation found another magnetic state stabilized near $n=1$: 
a layer-type antiferromagnetic state with a magnetic vector 
$\vec q = (\pi,0,0)$.\cite{Ulmke98} 
The semiclassical approximation is successful in finding 
the layer-type antiferromagnetic state near $n=1$. 
Computed N{\'e}el temperature at $n=1$ is $T_N \sim 0.048$ which agrees with 
the QMC result within the statistical error of QMC. 
Better agreement in the N{\'e}el temperature than in the Curie temperature 
may be attributed to the large $U$ used in this calculation and to the suppression of the
charge fluctuations near half-filling. 
According to the two-site DMFT, critical $U_c$ to the Mott transition at half filling is 
estimated to be $U_c = 6 v$ 
(correct value is expected to be slightly smaller then this).\cite{Potthoff01} 
With the parameter used, $U=6$, the system is in a Mott insulating state at $n=1$, 
and the charge fluctuation is suppressed. 
Therefore, the thermal fluctuation of spin field dominates the magnetic transition.  
In contrast to QMC, ferromagnetic and antiferromagnetic states contact with each other in 
the semiclassical approximation. 
(In the semiclassical result, $T_C$ and $T_N$ indicate the instability of 
the ferromagnetic and antiferromagnetic states, respectively, to the paramagnetic state.)
In the light of the overestimation of upper critical density $n_c$ for the ferromagnetism,
this failure is also supposed to be from the neglect of the quantum fluctuation of HS field. 
This point remains to be addressed in the future work. 

In the metallic region, one needs to fix the chemical potential according to the density-- 
this must be done at each interaction strength and temperature, which is time consuming. 
Even in this case, the semiclassical scheme is found to be computationally very cheap. 
Typical CPU time is less than 5~mins. at $U=2$ and $T=0.05$ using a commercial PC. 

Summarizing this section, we investigated the magnetic behavior of 
the Hubbard model on infinite- and three-dimensional FCC lattices 
using the semiclassical approximation. 
The magnetic phase diagram computed by the semiclassical approximation show 
reasonable agreement with QMC. 
The ferromagnetic Curie temperature in metallic region is found to be 
within a factor of $\sim 2$ compared with QMC, 
and the antiferromagnetic N{\'e}el temperature near $n=1$ 
is found to agree with QMC within statistical error. 
Better agreement between the semiclassical approximation and QMC in the N{\'e}el temperature 
near integer-filling is supposed to be from the fact that the charge fluctuation
is suppressed by the correlation, so that thermal spin fluctuations dominate the transition 
as in the half-filled square-lattice case. 
Poorer agreement in the Curie temperature is expected to be from the neglect of 
quantum fluctuation or charge fluctuation.  

\section{Application of semiclassical approximation to DCA and fictive-impurity method}

Our semiclassical approximation is easily combined with cluster schemes, 
such as dynamical cluster approximation\cite{Hettler98} (DCA) and 
real-site cluster extension of DMFT.\cite{Kotliar01,Biroli02,Okamoto03} 
As a simplest example, 
we use the two-site DCA and fictive-impurity\cite{Okamoto03} (FI) methods
to study the Hubbard model on a square lattice. 

We consider half-filling, thus charge fields are absorbed into the chemical potential shift. 
The Weiss field becomes $2 \times 2$ matrix $\bf a$, and in the paramagnetic phase, 
this has a form 
\begin{equation}
{\bf a} = a_0 {\bf 1} + a_1 \mbox{\boldmath $\tau$}_x , 
\end{equation}
with $a_0$ and $a_1$ being the on-site and NN Weiss fields, respectively. 
$\bf 1$ and $\mbox{\boldmath $\tau$}_x$ are the $2\times2$ matrices acting on 
orbital (impurity site) space. 
The HS transformation is performed at each impurity sites $i=1,2$ with spin field $\varphi_i$.  
The effective potential for HS field becomes 
\begin{eqnarray}
V(\varphi_1,\varphi_2) \!\! &=& \!\! \frac{1}{4U} \bigl( \varphi_1^2 + \varphi_2^2 \bigr) 
\nonumber \\
&& \!\! - T \, {\rm Tr} \, {\rm ln} \biggl[-{\bf a} 
- \frac{1}{2} 
\biggl(
\begin{array}{cc} 
\varphi_1& \\ 
& \varphi_2 
\end{array}
\biggr)
\mbox{\boldmath $\sigma$}_z \biggr]. 
\label{eq:Veff2}
\end{eqnarray}
Here, $\mbox{\boldmath $\sigma$}_z$ is acting on spin space, and 
Tr is taken over Matsubara-frequency, orbital and spin indices. 
Then, the partition function is given as 
\begin{equation}
Z_{approx} = \int d \varphi_1 d \varphi_2 \exp \{ -\beta V (\varphi_1,\varphi_2) \}. 
\end{equation}
Finally, on-site and NN-site Green functions, $G_0$ and $G_1$, respectively, are obtained via
\begin{eqnarray}
G_0 = \frac{1}{2} \frac{\delta \ln Z_{approx}}{\delta a_0}, \\
G_1 = \frac{1}{2} \frac{\delta \ln Z_{approx}}{\delta a_1}. 
\end{eqnarray}
The self-consistency equations for DCA are closed following the scheme presented in 
Ref.~\onlinecite{Hettler98}. 
The self-consistency equations for FI method and the general formalism for 
the cluster extension of DMFT are given in Ref.~\onlinecite{Okamoto03}. 

\begin{figure}%[htbp]
\begin{center}
\includegraphics[width=0.8\columnwidth,clip]{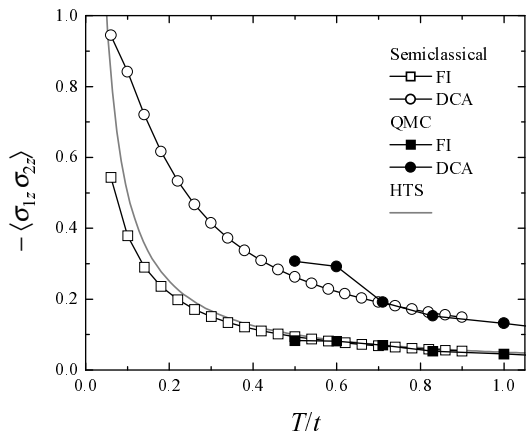}
\end{center}
\caption{Nearest-neighbor spin correlations $- \langle \sigma_{1z} \sigma_{2z} \rangle$
of a square-lattice Hubbard model ($U/t=20$) as functions of $T$ 
computed using two-site DCA (open circles) and the FI method (open squares), 
compared to QMC (filled symbols) and high-temperature series expansion(light line).
}
\label{fig:SScomp}
\end{figure}

Nearest-neighbor spin correlations $-\langle \sigma_{1z} \sigma_{2z} \rangle$ 
computed by DCA and FI with the semiclassical approximation and QMC 
as functions of temperature for $U=20t$ are shown in Fig.~\ref{fig:SScomp}. 
Similar spin correlation obtained by QMC in both DCA and FI supports the applicability of 
the semiclassical approximation to the cluster DMFT. 
For comparison, the same quantity computed using a high-temperature series expansion (HTS) 
for the $S=1/2$ NN Heisenberg model with $J=t^2/U$ are also shown. 
It is seen that spin correlation obtained from the DCA is much larger than the HTS result. 
We believe the origin of the discrepancy between DCA and HTS is that DA is equivalent to imposing a 
periodic boundary condition in all the directions in a real-space cluster is adopted. 
Thus, in the two-site DCA, one has a model in which two sites are connected via $z$ bonds 
[connectivity $z=4$ in a square lattice]
and the spin correlation is thus overestimated. 
Interestingly, FI method gives almost identical curves to HTS. 
(slight deviation can be seen below $T \sim 0.4t$.) 
A detailed study of multi-site cluster models using semiclassical methods and QMC 
will be presented elsewhere.\cite{Fuhrmann05} 

\section{Equilibration and partition phase space}

In this section, we point out an additional advantage of the semiclassical approximation. 
A key issue in Monte-Carlo simulations is equilibration, which is particularly difficult in systems 
in which the phase space is partitioned into several nearly equivalent minima, separated by large barriers. 
Local update techniques require extremely long runs to climb over barriers, 
while global update techniques are expensive and sometimes inconvenient 
to implement and bring their own convergence issues. 
The partitioned phase space phenomenon occurs frequently in strongly-correlated models, and represents a
significant obstacle to practical computations. 
The semiclassical approximation, by contrast, is inexpensive enough that the entire (semiclassical)
phase space can be sampled. 

\begin{figure}%[htbp]
\begin{center}
\includegraphics[width=0.8\columnwidth,clip]{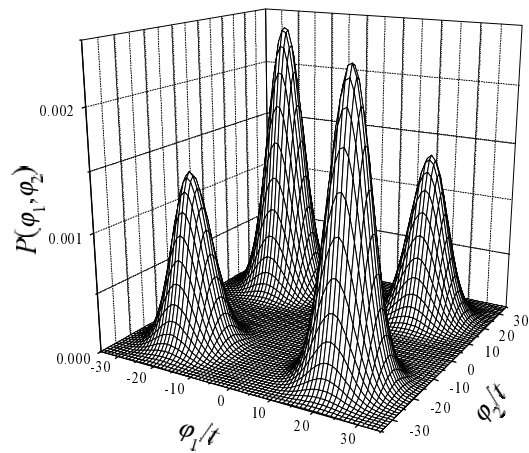}
\end{center}
\caption{Distribution of the spin fields $\varphi_1$ and $\varphi_2$, $P(\varphi_1,\varphi_2)$,
computed by two-impurity DCA for the Hubbard model on a square lattice. 
$P(\varphi_1,\varphi_2)$ is 
defined by $P(\varphi_1,\varphi_2) = \exp \{-\beta V(\varphi_1,\varphi_2)\}/Z_{approx}$. 
Parameters are $U/t=20$ and $T/t=0.5$. 
Larger peaks at $(\varphi_1,\varphi_2) \simeq (\pm 20t, \mp 20t)$ 
than $(\varphi_1,\varphi_2) \simeq (\pm 20t, \pm 20t)$ indicate 
antiferromagnetic correlation.}
\label{fig:P_DCA}
\end{figure}

For example, Fig.~\ref{fig:P_DCA} shows the distribution function of spin HS fields 
for the two-impurity DCA for the square-lattice half-filled Hubbard model
defined by $P(\varphi_1,\varphi_2) = \exp \{-\beta V(\varphi_1,\varphi_2) \}/Z_{approx}$ computed 
at temperature $T/t=0.5 > T_N$ (see Fig.~\ref{fig:TN_square}). 
The partitioning of phase space is evident. 
It is seen that the distribution has twofold symmetry, not fourfold, in $\varphi_1-\varphi_2$ plane; 
$\varphi_1>\varphi_2$ and $\varphi_1<\varphi_2$ are not equivalent. 
Larger peaks at $(\varphi_1,\varphi_2) \simeq (\pm 20t, \mp 20t)$ than 
$(\varphi_1,\varphi_2) \simeq (\pm 20t, \pm 20t)$ indicate that the antiferromagnetic correlations 
prevail far above the N{\'e}el temperature, giving the result shown in Fig.~\ref{fig:SScomp}. 
Reliable estimates of $-\langle \sigma_{1z} \sigma_{2z} \rangle$ requires sampling of all four extreme, 
with correct weights, which is very difficult to achieve in QMC at low $T$ 
(this is why we do not present data at $T<0.5t$, and statistical errors are evident even at $T=0.6t$). 
However, the integrals involved in the semiclassical method may be performed with no difficulty. 

\section{Summary and discussion}

In this paper, we investigated the semiclassical approximation to 
the continuous Hubbard-Stratonovich transformation as an impurity solver 
of DMFT method for the correlated-electron models. 
The Hubbard-Stratonovich transformation introduces two auxiliary fields, coupling to 
the electron spin and charge density respectively. The semiclassical approximation consists of 
retaining only the classical (zero-Matsubara-frequency) component in the functional integral 
over the spin field, 
and, for each value of the spin field, using a steepest descents approximation to approximate 
the integral over the charge field by the value which extremizes the action at the given value
of the spin field [see Eq.~(\ref{eq:xi})]. 
This treatment of the charge field was found to be essential for achieving reasonable results: 
see Fig.~\ref{fig:DOS_xi} and the associated discussion. 
The semiclassical approximation captures the thermal fluctuation of spin field efficiently 
beyond harmonic (Gaussian fluctuation) approximation and 
is applicable to both the metallic and insulating region. 
Estimates obtained using the Gaussian fluctuation approximation indicate that the semiclassical 
approximation gives reasonable results at larger $U$ (when spin fluctuations are soft) or at 
high temperature. (Sec.~III)
We applied the semiclassical approximation to the single-band Hubbard model 
on a two-dimensional square lattice (half-filling) (Sec.~IV) and 
infinite- and three-dimensional FCC lattices (finite doping) (Sec.~V). 
The semiclassical approximation is found to give reasonable results, with accuracy improving for 
stronger couplings and in situations where charge fluctuations are suppressed. 
A particularly attractive finding is that the procedure finds multiple phases when these exist.
(Fig.~\ref{fig:TC_fcc_3d}) 
The key point appears to be that, at stronger correlations, the physically important fluctuations 
become very soft (justifying a semiclassical approximation) but very anharmonic 
(requiring an integral over all field configurations). 
Due to the neglect of quantum effects, the semiclassical approximation fails to 
reproduce correctly the quasiparticle resonance, which will certainly be important at weak coupling 
or at low temperature. 
One possible improvement of this failure would be using a interpolative method\cite{Savrasov04}
to reproduce the low-energy quasiparticle in a paramagnetic metallic region. 
For the half-filled square lattice, 
we showed the semiclassical approximation gives reasonable 
behavior of the self-energy which is comparable to QMC in a wide range of parameters. 
The density of states and N{\'e}el temperature computed by the semiclassical approximation 
are found to be in very good agreement with QMC. 
In the metallic FCC lattice, 
the DOS has a two-peak structure corresponding to the lower- and upper-Hubbard bands 
in a paramagnetic phase. These structures evolve into majority- and minority-spin bands 
as the temperature is decreased through the ferromagnetic transition. 
Comparison of the DOS computed by the semiclassical approximation and by QMC shows 
reasonable agreement, especially as concerns the occupied state. 
Ferromagnetic Curie temperatures computed by the present method are found to be 
in the range of factor of 2 compared with the QMC. 
$T=0$ phase boundaries are found to be within the error of 20~\%. 
The semiclassical approximation for the Curie temperature shows better agreement than two-site DMFT, 
and $T=0$ phase boundaries are found to be in the same range of two-site DMFT. 
As for the three-dimensional FCC lattice, the semiclassical approximation is able to 
detect antiferromagnetic state near the integer filling consistent with QMC. 
N{\'e}el temperature in this case agrees with QMC within the statistic error. 
In Sec.~VI, we apply the semiclassical approximation to two-impurity DCA and 
real-space cluster DMFT (fictive-impurity) for the square-lattice Hubbard model at half-filling, 
and investigate how spatial correlation is taken into account in the present approximation.  
We confirmed that the short-range (nearest-neighbor) spin correlation prevails 
above the N{\'e}el temperature. 
Comparison of nearest-neighbor spin correlation between the semiclassical approximation 
and QMC shows good agreement in ranges where the QMC calculations are well converged. 
In Sec.~VII, equilibration problem in QMC associated with the partition of phase is discussed. 
This problem does not occur in the semiclassical approximation. 

In conclusion, we mention some open scientific questions which can be addressed by 
the semiclassical method, and mention two areas in which further investigation and improvement 
would be desirable. 
A key opportunity involves systems with several sites and/or several orbitals. 
In such cases, QMC suffers from ``sign'' problems, and both QMC and ED method 
scale poorly with system size. 
The semiclassical method does not suffer from ``sign'' problem; while 
there appeared similar problem associated with the imaginary coefficient for the charge-field, 
this is resolved by applying the saddle-point approximation to the charge-field. 
When it is applied to systems with $N>1$ orbitals (on one or more sites), 
computational time for the semiclassical approximation 
scales of $P_\varphi^N$ with $P_\varphi$ and $N$ being a total number of descretized HS field 
and total number of sites/orbitals, respectively. 
By contrast in Monte-Carlo methods, the scaling is $[(\beta U)^2 N_{MC}]^N$ 
($N_{MC}$ is a number of Monte-Carlo sampling, typically $N_{MC} \sim 10^5$), 
and by $4^{N (1+N_{bath})}$ for ED 
($N_{bath}$ is a number of ``bath'' sites per impurity orbital, and 
$N_{bath}$ should be larger than 2--3 from the overestimation of $T_C$ by two-site DMFT). 
As can be seen in the Fig.~\ref{fig:P_DCA}, dominant contribution is known to be from 
$|\varphi| \sim U$, and if need it would be possible to simplify the $N$-dimensional integral 
over HS fields. 
Thus, the semiclassical approximation appears to be an attractive option for studying  
larger clusters than QMC.

As for the multisite problem, 
we have only applied the semiclassical approximation to the single-orbital Hubbard model 
at half-filling with DCA and FI method. 
In this case, the charge-field is absorbed into the chemical potential. 
Away from the half-filling, one needs to fix the charge-field at each configuration of 
the spin-fields according to Eq.~(\ref{eq:xi}) generalized to the multisite situation. 
There is no proof that Eq.~(\ref{eq:xi}) has unique solution in this situation 
although we have so far not encountered problems. This issue needs further investigation. 

As one of the applications of the semiclassical approximation, 
investigation of the systems with degenerate orbitals like transition-metals oxides with 
$e_g$ or $t_{2g}$ electrons would be interesting.\cite{Imada98,Tokura00} 
Theoretical studies of the ferromagnetism in these situation have been presented 
(Refs.~\onlinecite{Tyagi75,Hasegawa83,Kakehashi86}), 
but these works dealt with spin-fluctuations in the metallic state, 
fluctuation and ordering associated with the quadrupole moment for 
$e_g$ and/or $t_{2g}$ orbital was neglected. 
Hubbard-Stratonivich transformation including spin and quadrupole moment for two-band Hubbard 
interaction for $e_g$ orbital has been introduced in Ref.~\onlinecite{Ishihara97}. 
%However, these calculations had been based on a simplified interaction which breaks 
%the orbital symmetry. 
However, pin and orbital transition at finite temperature using the multi-band Hubbard model 
with full symmetry\cite{Mizokawa95} remains to be investigated. 

Another promising application of the semiclassical approximation would be 
the spatially inhomogeneous systems, where the computational expense associated with other aspects of
the problem renders an inexpensive impurity solver essential. 
Recently, two of the authors (S.O. and A.J.M.) applied two-site DMFT to 
the spatially-inhomogeneous heterostructure problem and investigated the evolution 
of the electronic state (quasiparticle band and upper- and lower-Hubbard bands) 
as a function of distance from the interface.\cite{Okamoto04} 
Applying the semiclassical approximation to such systems 
to investigate the possible spin (and orbital) orderings\cite{Nature} 
is an interesting and urgent task.  

\acknowledgements
The authors acknowledge useful discussions with B. G. Kotliar and M. Potthoff. 
This research was supported by the JSPS (S.O.) 
and the DOE under Grant ER 46169 (A.J.M.).

\end{document}